\documentclass[a4paper,12pt]{article}
\usepackage[utf8]{inputenc}
\usepackage{amssymb}
\usepackage{amsmath}
\usepackage{mathrsfs}
\providecommand{\pp}[2][]{\frac{\partial #1}{\partial #2}}
\title{Stochastic Quantization of Axial Vector Gauge Theories with Fermions}
\author{A. K. Kapoor\footnote{akkapoor@cmi.ac.in;akkhcu@gmail.com}\\ Chennai 
Mathematical 
Institute\\ H1, SIPCOT IT Park, Siruseri, \\ Kelambakkam, Tamil Nadu 603103, 
INDIA}

\begin{document}

\maketitle

\begin{abstract}
The stochastic quantization scheme proposed by Parisi and Wu in 1981 is known
to have differences from conventional quantum field theory in higher orders.
It has been suggested  that some of these new features  might give rise to a
mechanism  to  explain tiny fermion masses as arising due to radiative
corrections. In view of importance for need of going beyond the standard model, 
in this article some features  of U(1) axial vector gauge theory in Parisi 
Wu stochastic quantization scheme are reported. Renormalizability of  a massive 
axial vector gague theory coupled  to a massless fermion appears as one of the
important conclusions.
\end{abstract}
\baselineskip=16pt
In an earlier article \cite{akk15}, it has been suggested that fermion masses
may get generated as radiative corrections in theories with \(U(1)\) axial 
vector gauge field when Parisi Wu stochastic quantization method (SQM) is
used\cite{SQM1}. It may be recalled that originally SQM was proposed as a scheme
to quantize gauge theories without gauge fixing. It has several important
features  which do not have an analogue in conventional quantum field theory
(CQFT). One such  well known feature is Zwanziger stochastic gauge fixing. We
refer the reader to \cite{SQM2,SQM3,SQM4} for review of Parisi Wu quantization.
For references to earlier works on renormalization of fermionic theories in SQM
can be found in \cite{Sand}.

In this letter  I will be concerned primarily with a four dimensional  axial
vector \(U(1)\) gauge theory coupled to a massless fermion. Features that 
appear in non-abelian vector and axial vector gauge theories will be 
briefly addressed in the end. 

The Parisi Wu  scheme is formulated in terms of the Euclidean action.
The {\it Euclidean action} for corresponding  four dimensional CQFT of a 
\(U(1)\) axial vector gauge theory will be taken to be  
\begin{eqnarray}
   S_E &=& \int d^4 x \mathscr{L} d^4x \\
   \mathscr{L} &=&  \bar{\psi}(-i\gamma_\mu D_\mu + m)
\psi - \frac{1}{4} F_{\mu\nu}F_{\mu\nu} + \frac{M^2}{2}A_\mu A_\mu  -
\frac{1}{2\alpha}(\partial\cdot A)^2+ \mathscr{L}^\prime
\end{eqnarray}
where
\begin{eqnarray}
 D_\mu = \partial_\mu -ig\gamma^5 A_\mu,\qquad F_{\mu\nu} = \partial_\mu A_\nu-
\partial_\nu A_\mu,
\end{eqnarray}
and $\mathscr{L}^\prime$ is part of the Euclidean Lagrangian describing other 
fields
which may be present.

In the following we will take mass of the fermion to  be zero, \(m=0\). A term
\(\frac{1}{2\alpha}(\partial_\mu A_\mu)^2\) has been included so that the
degree of divergence remains bounded. We will obtain Ward identities which
ensure that unphysical degree of freedom, the scalar component of the
axial vector  field, decouples. In fact the scalar part of the gauge field
will  turn out to be a free field.

In the stochastic quantization scheme the fields are stochastic processes 
obeying Langevin equations in fictitious time. We will use  $x$ to 
collectively denote  all four components of four vector 
$x_\mu=(x_1,x_2,x_3,x_4)$ and $t$ will be used to denote the fictitious time 
or the stochastic time.

In the equilibrium limit the SQM  
equal time correlation functions of the fields coincide with the Green 
functions of Euclidean CQFT. The basic Langevin equations of SQM are given by
\begin{eqnarray}
  \frac{\partial A_\mu(x,t)}{\partial t}
  &=& - \gamma^{-1} \frac{\delta S_E}{\delta A_\mu} +
\eta_\mu(x,t)\label{EQ02}\\
 \frac{\partial \psi(x,t)}{\partial t}
 &=& - \int dx^\prime K(x,x^\prime) \frac{\delta S_E}{\delta
\bar{\psi}(x^\prime)} + \theta(x,t)\label{EQ03} \\
\frac{\partial \bar{\psi}(x,t)}{\partial t}
 &=&  \int dx^\prime  \frac{\delta S_E}{\delta {\psi}(x^\prime)} K(x,x^\prime)
+ \bar{\theta}(x,t)\label{EQ04}.
\end{eqnarray}
 The Gaussian white noises
$\eta_\mu(x,t), \theta(x,t), \bar{\theta}(x,t)$ are assumed to have averages
\begin{eqnarray}
  \langle\eta_\mu(x,t)\eta_\nu(x^\prime,t^\prime) \rangle =
2\gamma^{-1}\delta_{\mu\nu}\delta(x-x^\prime)\delta(t-t^\prime) \label{EQ05}\\
  \langle\theta(x,t)\bar{\theta}(x^\prime,t^\prime) \rangle =
2 K(x,x^\prime)\delta(t-t^\prime) .\label{EQ06}
\end{eqnarray}
In an operator formalism proposed by Namiki  et al. \cite{SQM3}, the SQM is
formulated as a field theory in five dimensions following a definite
prescriptions of writing the action in five dimension. A stochastic momentum
field is introduced for each field in the Lagrangian. Letting \(\pi_\mu,
\bar{\omega},\omega\) to denote the stochastic momentum corresponding to
the gauge field \(A_\mu\) and the fermionic fields \(\psi, \bar{\psi}\),
the stochastic action \(\Lambda\) of the five dimensional field theory takes
the following form.
\begin{equation}
 \Lambda = \int dx dt\left( \pi_\mu \frac{\partial A_\mu}{\partial
t}  + \frac{\partial\bar{\psi}}{\partial t}\omega  + \bar{\omega}
\frac{\partial\psi}{\partial t} - {\mathcal H}.
 \right) \label{EQ07}
\end{equation}
where
\begin{eqnarray}
{\mathcal H}&=& \left[ \gamma^{-1} \pi_\mu\pi_\mu + 2\bar{\omega}K \omega
-\bar{\omega}\tilde{K}\frac{\delta S_E}{\delta \bar{\psi} } + \frac{\delta
S_E}{\delta \psi}\tilde{K} \omega -
\gamma^{-1}
\pi_\mu\frac{\delta S_E}{\delta A_\mu}
\right] \label{EQ08}.
\end{eqnarray}
Here \(K(x,t)\) is a suitable kernel that  needs to be used for the fermions.
For our present discussion an explicit expression of the kernel for fermions
is not required.
We write down expressions for two point function for the axial vector fields
\begin{eqnarray}\label{EQ91}
  \big\langle T[A_\mu(k,t) \pi_\nu(k', t') \big\rangle &=& \theta(t-t')
     \delta(k+k') P_{\mu\nu}  e^{-(k^2+m^2)(t-t')}\label{EQ92}\\
    \big\langle T[A_\mu(k,t) A_\nu(k', t')] \big\rangle &=& \delta(k+k')
P_{\mu\nu}   e^{-(k^2+m^2)|t-t'|}\\
   \big\langle T[\pi_\mu(x,t) \pi_\nu(k',t')]\big\rangle &=& 0.\label{EQ93}
\end{eqnarray}
where
\begin{equation}
  P_{\mu\nu} = \frac{\delta_{\mu\nu}}{k^2+M^2} - \frac{k_\mu
k_\nu}{M^2}\Big( \frac{1}{k^2+M^2} - \frac{k_\mu k_\nu}{k^2+\alpha M^2}\Big)
\end{equation}

The two point correlation function, such as the one appearing in \eqref{EQ91}, 
is also known as \(G\) propagator for the axial vector field. This is just the 
Green function for the Langevin equation. The presence of \(\theta\) function 
in the \(G\)- propagator  be noted.

Recall that in CQFT of Yukawa coupling \(g\bar{\psi}\gamma^5\psi\) of a spin
zero field \(\phi\) coupled to a fermion the scalar four point function is
divergent and a \(\phi^4\) counter term is required. Even if such a term is not
included in the Lagrangian as coupling term, it needs to be added to the
Lagrangian in order to remove ultraviolet divergences. Therefore, the
renormalized theory has a new unknown parameter, \(\lambda\), appearing in
the renormalized Green functions. Something similar happens in SQM of axial 
vector \(U(1)\) gauge theory we are considering here, but  there are important 
differences.

The three point vertex function \(\pi_\mu,  A_\nu, A_\sigma\) is linearly
divergent and a counter term needs to be  added to the stochastic Hamiltonian.
The counter term has the form \(f_{AA\pi}\epsilon_{\mu\nu\lambda\sigma}
 \pi_\mu(x,t)A_\nu(x,t)\partial_\sigma A_\lambda(x,t)\). It must be
mentioned here that this counter term is gauge invariant. This is easily checked
by noting that  \(\pi_\mu\) does not  transform under gauge transformations. The
renormalized theory has an extra parameter, the coefficient \(f_{AA\pi}\) of the
new \(A\)-\(A\)-\(\pi\) counter term.

At this point we mention two important features of SQM of axial gauge theory.
For any choice of kernel the stochastic action breaks the chiral symmetry at
the tree level itself. The presence of the new counter term mentioned mentioned
here breaks the stochastic supersymmetry responsible for equivalence of SQM and
CQFT formalism of the model. These two departures from expectations based on
experience with other models, such as scalar fields, formed the basis of
suggestion that fermion masses could be generated fully by radiative
corrections and a possible explanation may be found for tiny  masses of
leptons.

Coming back to the \(\pi\)-\(A\)-\(A\) counter term, note that this will
contribute a linear divergence to the three point Green
function \(A_\mu(y)A_\nu(y)A_\lambda(z)\) in the equilibrium limit. A closure
examination reveals that this three point function is completely fixed by  the
Bose symmetry and thus the linear divergence can be eliminated unambiguously
\cite{AKK79}.
Based on the known non-renormalization property of axial anomaly, it is expected
that in SQM \(f_{AA\pi}\) will not get renormalized as perturbation terms of
orders higher than three are considered.

If one sets \(f_{AA\pi}\)  to zero, we will be back to anomalous theory
equivalent to CQFT. If one sets it to a particular value, that value will not
receive any corrections due to renormalization.  Hence it can be as tiny as we
wish without need for assigning any more theoretical explanation.

The actual value  of \(f_{AA\pi}\) coupling constant  will have to be fixed from
experimental data or some other theoretical requirement.  If correct number of
fermions are  present to ensure anomaly cancellation the model will be
renormalizable and one can set \(f_{AA\pi}\)  to zero.

We will now argue that in models where there is no cancellation of anomaly, we
can choose value of \(f_{AA\pi}\) so as to enforce  conserved axial vector
current Ward  identities in the limit of fermion masses are taken to be
zero. The relevant Ward identities in  presence of a \(\pi\)-\(A\)-\(A\) term
follow
quite trivially from the equations of motion of  the stochastic
momentum field \(\pi_\mu\) and taking equal time limit.

Let \(\Xi\) denote an arbitrary  product of fields \(A_\mu(x_k,t_k)
,\bar{\psi}(y_\ell,t_\ell), \psi(z_m, t_m)\). Then the equation of motion of
\(\pi_\alpha(x,t)\) can be written down by considering the generating
function of the correlation functions as a functional integral and setting its
variation w.r.t. \(\pi_\mu(x,t)\) equal to zero.
\begin{equation}
  \Big\langle \frac{\delta \Lambda}{\delta \pi_\mu(x,t)} \Xi  \Big\rangle
 =  \Big\langle \frac{\delta}{\delta \pi_\mu(x,t)} \Xi 
\Big\rangle=0 \label{EQ10}
\end{equation}
Here  \(\langle \cdots\rangle\) denotes the \(n\)- point functions of the five
dimensional theory. It is easy to see that
\begin{equation}\label{EQ11}
  \frac{\delta \Lambda }{\delta \pi_\mu(x,t)} =
\pp[A_\mu(x,t)]{t}-\frac{2}{\gamma}\pi_\mu(x,t) + \frac{\delta S}{\delta
A_\mu(x.,t)}
\end{equation}
As a first step to get the desired  Ward identity we substitute \eqref{EQ11} in
\eqref{EQ10} and take equal time limit of the above equation. In the
equilibrium limit the equal time correlation functions of SQM give the Green 
functions of CQFT. For equal time correlation functions, the second 
term involving the correlation function of \(\pi_\mu(x,t)\) becomes zero 
when considering only connected diagrams.
This can be seen as a consequence of the following facts.
\begin{enumerate}
  \item The \(\pi_\mu\)-\(\pi_\nu\) two point correlation function is zero.
  \item Every interaction term contains one power of  stochastic momentum.
  \item Due to presence of theta  function, \(G\) lines cannot form a loop
        because integration region over times corresponding to internal
        vertices shrinks to a point.
  \item At least one external \(G\) line starting from an external vertex will
        be  continuously connected by a chain of \(G\) lines to some exteral
        vertex. At equal times, this contribution drops out if the chain passes
        through an internal vertex. This is again due to the fact that the
        time integration corresponding to the internal vertices on \(G\) line
        collapses to zero,
  \item From the above properties it follows that at equal times
contribution comes from only those (disconnected) diagrams in which external
\(\pi_\mu\) line is connected to  some external line \(A_\nu\). At equal times
the \(G\) propagator 	reduces to \((1/2)\delta(x-y)\):
\begin{equation}
  \big\langle T[A_\mu(x,t)\pi_\nu(y,t)]\big\rangle_0 = \frac{1}{2} 
\delta(x-y)\delta_{\mu\nu}.
\end{equation}
\end{enumerate}
Here \(\langle...\rangle_0\) denotes equal time correlation function.
Thus the equation of motion for the \(\pi_\mu\) field takes the form.
\begin{equation}
  \big\langle T \frac{\delta S_E}{\delta A_\mu}\Xi \big\rangle_0 -  \big\langle
T \frac{\delta}{\delta A_\mu}\Xi \big\rangle_0 = 0
\end{equation}
Now let us  take the divergence
\begin{equation}
   \big\langle T \left(\partial_\mu  \frac{\delta S_E}{\delta
A_\mu}\right)\Xi \big\rangle_0 - 
 \big\langle T
\partial_\mu \frac{\delta }{\delta A_\mu}\Xi \big\rangle_0 = 0
\end{equation}
and write out all terms explicitly. This leads to the following Ward identity.
\begin{equation} 
  \Big\langle T \big[(-\Box + \alpha^2M^2) (\partial\cdot A) + \partial_\mu 
J^5_\mu
\big] \Xi \Big\rangle_0 =\Big\langle T
\partial_\mu \frac{\delta }{\delta A_\mu}\Xi\Big\rangle_0  \label{EQ12}
\end{equation}
where \(J^5_\mu\) is the axial vector current of the fermion and its divergence
is of the form \(\epsilon_{\mu\nu\lambda\sigma} F_{\mu\nu}F_{\lambda\sigma}\) 
and is nonzero due to the presence of anomaly.\footnote{The axial vector
anomaly has been discussed in SQM by many authors, see sec 2.1 of Namki et al.
\cite{SQM2} }
When a \(\pi\)-\(A\)-\(A\) counter term is included, \eqref{EQ12} gets modified
and the
new Ward identity becomes
\begin{eqnarray}
\lefteqn{  \Big\langle T \big[-\Box + \alpha^2M^2) (\partial\cdot A) +
\partial_\mu
J^5_\mu \Big] \Xi\Big\rangle}&&\nonumber\\
 + && \Big\langle T\big[f_{AA\pi}\epsilon_{\mu\nu\lambda\sigma}
 \partial_\mu A_\nu(x,t)\partial_\sigma A_\lambda(x,t)\Xi \big] \Big\rangle_0 
=\Big\langle T
\partial_\mu \frac{\delta }{\delta A_\mu}\Xi\Big\rangle_0 . \label{EQ22}
\end{eqnarray}
It is obvious that the structure of the term proportional to \(f_{AA\pi}\) is
such that it will cancel the anomalous divergence of the fermion current. In
presence of this term, the current coupled to the axial vector field has
an extra term proportional to \(f_{AA\pi}\). {\it The structure of  the extra
term ensures that a choice  of \(f_{AA\pi}\) exists such that the current
coupled to the axial gauge field is conserved.}

The Ward identity \eqref{EQ22}, with suitable choice of \(f_{AA\pi}\) 
implies that the unphysical scalar  part of the axial vector decouples from the 
physical sector. This can be seen by amputating the external lines and putting 
them on mass shell and noting that the resulting equation
implies that \(\partial\cdot A\) satisfies free field equation.

{\it This leads to the result that this theory is renormalizable and
physically acceptable when quantized using Parisi Wu stochastic quantization 
scheme.}

On the other hand if fermion multiplet is arranged to cancel anomaly, the
\(f_{AA\pi}\) counter term must be set equal to zero; a non-zero value will
mean that the current coupled to  the axial vector gauge field is not conserved
and the unphysical degree of freedom, \(\partial_\mu A^\mu\) may not decouple
and we may end up with a non-unitary theory.

It will be an interesting and useful exercise to see which of the above 
conclusions survive, and if any new results appear, in  a non-abelian gauge 
theory involving vector as well as axial vector gauge fields. It is 
straightforward to see that non-abelian version, of  \(\pi\)-\(A\)-\(A\)
counter term in \(U(1)\) theory, is \(\epsilon_{\mu\nu\lambda\sigma}D_\mu\pi_\nu
F_{\lambda \sigma}\). In SQM of  pure Yang Mills theory such term is gauge
invariant and renormalizable by power counting. A priori, there is no  reason
to exclude such
a term in SQM formulation of pure Yang Mills theory.

In the electroweak sector of  realistic model, inclusion of a
\(\epsilon_{\mu\nu\lambda\sigma} D_\mu\pi_\nu F_{\lambda
\sigma}\) term in case of Yang Mills field  will lead to
a wide variety of \(C,P,T\) violating  interactions depending on coupling to the
matter field. In the strong sector it will have important implications for
the \(U(1)\) problem and strong \(CP\) violation. A study of  phenomenological
consequences will have to wait until important issues for
a non-abelian gauge theories have been fully investigated.

I thank H.S. Mani for helpful discussions  and Harikumar a useful suggestion.

\end{document}